# TRAINING ONE MODEL TO DETECT HEART AND LUNG SOUND EVENTS FROM SINGLE POINT AUSCULTATIONS

*Leander Melms[1,7], Robert R. Ilesan[2*], Ulrich Koehler[3], Olaf Hildebrandt[3], Regina Conradt[3], Jens Eckstein[4], Cihan Atila[4], Sami Matrood[5], Bernhard Schieffer[6], Jürgen R. Schaefer[6], Tobias Müller[6], Julius Obergassel[7], Nadine Schlicker[1] and Martin C. Hirsch[1]*

[1]Institute of Artificial Intelligence, Philipps-University Marburg; [2]Clinic of Oral and Cranio-Maxillofacial Surgery, University Hospital Basel; [3]Division of Respiratory and Critical Care Medicine, Philipps-University, Marburg; [4] Division of Internal Medicine, University Hospital Basel, Basel, Switzerland; [5]Department of Gastroenterology, Endocrinology, Metabolism and Infectiology, Philipps-University, Marburg; [6]Cardiology Department, University Hospital Gießen and Marburg; Department of Internal Medicine; [7]Department of Cardiology, University Heart and Vascular Center Hamburg, Hamburg, Germany

\* Correspondence: robert.ilesan@unibas.ch

## ABSTRACT

**Objective:** This work proposes a semi-supervised training approach for detecting lung and heart sounds simultaneously with only one trained model and in invariance to the auscultation point. **Methods:** We use open-access data from the 2016 Physionet/CinC Challenge, the 2022 George Moody Challenge, and from the lung sound database HF_V1. We first train specialist single-task models using foreground ground truth (GT) labels from different auscultation databases to identify background sound events in the respective lung and heart auscultation databases. The pseudo-labels generated in this way were combined with the ground truth labels in a new training iteration, such that a new model was subsequently trained to detect foreground and background signals. Benchmark tests ensured that the newly trained model could detect both, lung, and heart sound events in different auscultation sites without regressing on the original task. We also established hand-validated labels for the respective background signal in heart and lung sound auscultations to evaluate the models. **Results:** In this work, we report for the first time results for i) a multi-class prediction for lung sound events and ii) for simultaneous detection of heart and lung sound events and achieve competitive results using only one model. The combined multi-task model regressed slightly in heart sound detection and gained significantly in lung sound detection accuracy with an overall macro F1 score of 39.2% over six classes, representing a 6.7% improvement over the single-task baseline models. **Conclusion/Significance:** To the best of our knowledge, this is the first approach developed to date for measuring heart and lung sound events invariant to both, the auscultation site and capturing device. Hence, our model is capable of performing lung and heart sound detection from any auscultation location.

## I. INTRODUCTION

Specific sounds are audible through auscultation elicited breathing and cardiac action, which are clinically classified based on their characteristics and occurrence. The essential sounds result from the rhythm of respiration (inhalation and exhalation) and from the cardiac action (first and second heart sounds) as well as from possible pathological organ disorders, such as for the lung auscultation - wheezing and crackles. In addition to informing that these auscultation events are present, and more importantly the temporal occurrence in conjunction with the frequency and duration data provide important information about the patient's status: The number of inspirations and the first heart sounds (as the beginning of the cardiac cycle) can be used to directly determine the vital signs of respiration and heart rate, as well as the respective rhythm. The duration data can also play a major informative role: A prolonged expiration, for example, may suggest airway obstruction. Furthermore, the temporal occurrence of cardiac and pulmonary events can be used to infer, for example, a respiratory arrhythmia. Recent work has also shown that the temporal occurrence of pathological lung sound events with respect to respiratory action may be indicative of various disease etiologies (1).

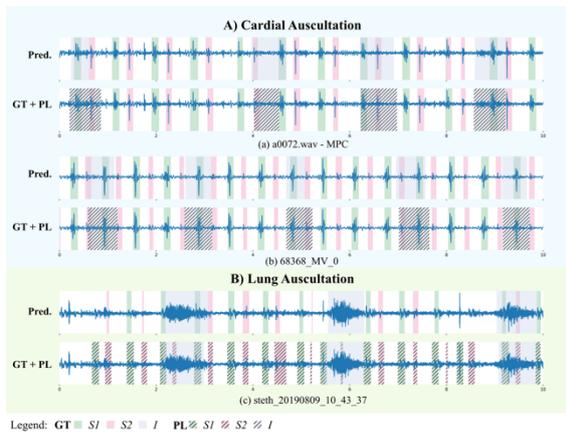

**Figure 1**. *Examples of automatically segmented **A)** heart sound and **B)** lung sound auscultation files (10 seconds) with ground truth (GT) and prediction labels (CRNN) of the PhysioNet/CinC Challenge 2016 dataset. Pseudo-labels for the background events are striped. GT = ground truth, PL = pseudo-label, $S_1$ = first heart sound, $S_2$ = second heart sound, I = inspiration*

Lung and heart sounds differ fundamentally in frequency characteristics, loudness, rhythm, occurrence, and duration of events. Rhythmic events such as heart sounds $S_1$/$S_2$ ($S_1$ marks the beginning of systole, $S_2$ marks the end of systole and beginning of diastole, Fig. 1) elicited by cardiac action, can always occur independently of the lung sound events. In lung auscultation, there are also tendential rhythmic events that follow a pattern, such as inhalation and exhalation. The lung sound events—crackles, wheezes, rhonchi, and stridor—can only occur within the inhalation and exhalation phases. If audible, an inhalation is physiologically always followed by an exhalation, just as the second heart sound, must always follow the first heart sound, if present and not superimposed.

All other sound events (crackles, wheezes, rhonchi, stridor) do not follow such sequencing. The specifics of the major heart and lung sound events can be summarized as follows:

- **$S_1/S_2$:** Short but repetitive events, with little variation in their acoustic features compared to lung sound events
- **Inspiration/Expiration**: Varying durations, varying sound continuities
- **Crackles**: Short time, instantaneous, discontinuous, explosive; heard during inspiration and sometimes during expiration; typically less than 20 ms in duration (1); influenced by the indifferentiable rubbing-sound noises; discontinuous, low or high-pitched; can occur during the inspiratory phase, expiratory phase, or both phases
- **Wheezing**: Short to long time, continuous (1), high-pitched; usually louder than the underlying breath sound; confusable with snoring, usually appear as high-intensity horizontal stripes on a spectrogram; can occur during the inspiratory phase, expiratory phase or both phases, continuous, high-pitched
- **Rhonchi**: Low-pitched, continuous musical; dominant frequency of 200 Hz or less; snoring quality (1)
- **Stridor**: Loud and high-pitched; caused by a turbulent flow through an obstruction of the upper airway; mainly inspiratory (1)

The temporal detection of audio events in an audio track with the specification of onset and offset times for a distinct number of target sound events is referred to as Sound Event Detection (SED). Within SED tasks, a further basic distinction is made between two degrees of complexity, whereby in simpler scenarios there is no temporal overlap of the individual different sound events, and in more complex tasks several different audio events take place simultaneously and thus overlap. The former variant is also called monophonic SED and the latter, due to the overlapping events, is polyphonic SED. Fig. 2 shows a segment of polyphonic SED labels, where the three events - inhalation, $S_1$, and exhalation - take place. Inhalation and $S_1$ overlap at time step t. Heart SED of the four events - $S_1$, systole, $S_2$, and diastole - is a monophonic SED task, as, by definition, the events cannot overlap. However, lung SED is always a polyphonic sound event detection task, as events happen simultaneously.

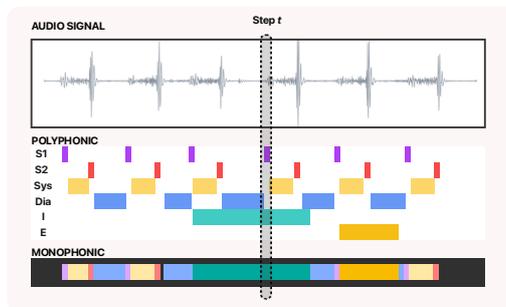

**Figure 2**: *Sound Events for the three event classes Inspiration, Expiration, and $S_1$ at different time steps. In polyphonic SED, sound events can overlap, as displayed at the time step t (black outlined rectangle).*

Auscultation of the heart and lungs is performed sequentially and separately during the physical examination. For example, auscultation of the heart is first started in the different projection areas of the heart valves on the patient's chest (see Fig. 3 A). Then the patient's back will be auscultated to better assess the lungs. Just like the separation of the sequence of cardiac and pulmonary auscultation, available open-source benchmark datasets are also separated between cardiac and pulmonary auscultation (see Fig. 3 B). In the datasets, either the cardiac or pulmonary events are annotated, although in many cases the respective other signal is also quietly perceptible in the background.

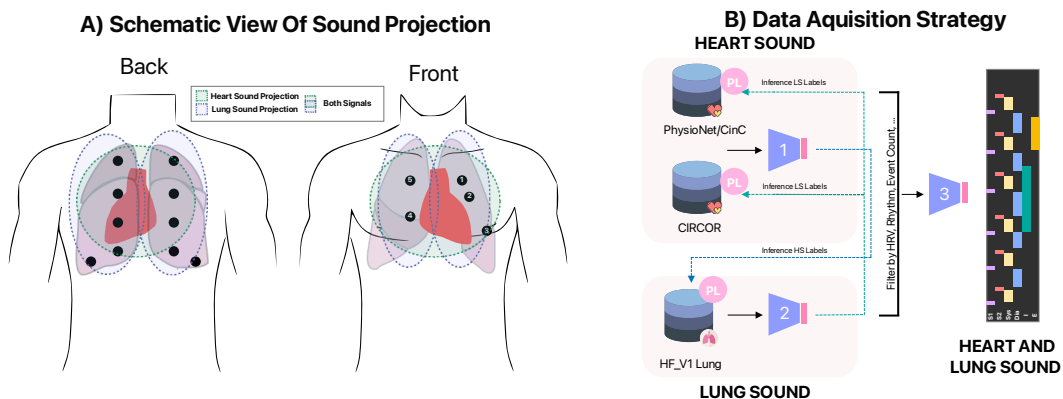

**Figure 3**: **A**: Auscultation points of the lungs and heart from the back and front. The preferred auscultation points for lung and heart auscultation are shown. The sound projections partially overlap. This is shown schematically with the colored dotted projection areas. 1 = Pulmonary valve, 2 = Erb's point, 3 = Mitral valve, 4 = Tricuspid valve, 5 = Aortic valve. **B**: *Proposed strategy for background signal label generation. First, two separate specialist models were trained to detect heart (1) and lung (2) sound events using ground-truth labels (GT). In the next step, these models were used to generate pseudo-labels (PL) for the corresponding other database. Then, GT and PL were merged to train a model (3) capable of detecting heart and lung sound events.*

In this work, we propose to detect both signals, regardless of auscultation location, using a single common model. The proposed novel training scheme for heart and lung SED is illustrated in Fig. 3. The model is based on the insights from semi-supervised learning that expressive information within the generated output of teacher-like models can be used to train a new robust model. In this way, all available heart and lung auscultation datasets can be used for training. The larger data variance should, i) make the model more robust, and ii) provide a better understanding of the event to be detected, regardless of whether it is heard in the background or foreground. In addition, due to the usage of only one model near real-time, inference on edge devices, such as smartphones, is feasible for continuous monitoring applications. To the best of our knowledge, this is the first approach developed to train an auscultation-site invariant end-to-end SED model for simultaneous heart and lung sound detection.

The main contributions of this work are:
1. We propose a semi-supervised inspired approach of generating hard pseudo labels (PL) for the background signal. For this, specialist models are trained separately on heart and lung SED and are subsequently used for inference (see Fig. 3).
2. We adapt semi-supervised training for the purpose of SED and investigate performance gains by incorporating large amounts of pseudo-labeled extra data.
3. We show that our newly trained combined multi-task model regresses only slightly in heart sound detection while gaining significantly in lung sound detection accuracy.

## II. RELATED WORK

### A. SED

**1) SED model architectures.** Convolutional Recurrent Neural Networks (CRNN), in which Convolutional Neural Network (CNN) layers precede temporal recurrent layers as feature extractors, are the dominant architecture for SED in competitions (2). They have the advantage of being able to connect expressive local information into a temporal context with the recurrent layers. Several successful works are based on such an architecture (3–5). In addition, stand-alone CNNs (6), recurrent neural networks (RNN) (7), and recently transformer-based networks (8,9) are also used in SED. Sound event models usually output a probability of being active at a given time step $t$ for each class. A (Mel) log spectrogram is often chosen as input, whereby, in order to preserve the time information, it is either not pooled or, for example, only pooled by half in the time axis in subsequent interpolation. Either softmax or sigmoid activation functions are used at the models' output layer for monophonic or sigmoid for polyphonic SED respectively. The model output is then often binarized by a threshold value to obtain a binary activity matrix.

**2) SED imbalance:** Data imbalance for SED concerns the distribution between the event classes, between active and inactive time frames and in time duration. To address the imbalance between rare and frequent events i.a. data augmentation (10–12), temporal subsampling (12), and few-shot learning (11) were investigated. Imoto et al. (13) were able to show that the differences in the distribution of active and inactive frames of the event classes can have a greater negative effect than that of class imbalance in SED. The authors proposed a new loss function Asymmetric Focal Loss (AFL) to reweight the losses, which addresses the class imbalance and active/inactive frame imbalance together. It takes the two arguments, $\gamma$ for balancing between classes and $\zeta$ for balancing between active/inactive frames. In their work, addressing both, the data imbalance between sound event classes ($\gamma$) and between active/inactive frames ($\zeta$), with AFL ($\gamma = 0.0625$, $\zeta = 1.0$) outperformed all other methods.

### B Heart sound detection

**1) State sequence.** In contrast to lung SED, heart sounds follow by definition a fixed state sequence ($S_1$-systole-$S_2$-diastole). Thereby, no overlapping events are possible and the event detection for heart sound detection is always monophonic. To preserve the physiological state order, complementary methods are often used in cardiac sound detection. One strategy is to force the output sequence of the model to allow only permissible transitions between the heart sound events (14). However, this method can be affected by noise, which leads to shifts in the heart sound events of $S_1$ and $S_2$. In other previous studies (14–16) the hidden semi-Markov model (HSMM) that introduces the state duration probability was used. In HSMM annotation is performed with an extended Viterbi algorithm, which combines the information of the whole signal for the sound event annotation. However, Messner et al (17) found that the extended Viterbi algorithm is error-prone in the presence of irregularities in the cardiac action due to, e.g. cardiac arrhythmias.

**2) Model architectures**: As in most SED challenges, CNN, RNN, Temporal Convolutional Networks (TCN), and CRNN architectures are mainly used for the heart SED. Messner et al. (17) used a bidirectional RNN with different inputs (41-bin log magnitude spectrograms, Mel Frequency Cepstral Coefficients (MFCCs) and Envelope features), achieving F1 scores for $S_1$ of 97.5% and 95.0% for $S_2$ on an independent test set. Renna et al. (14) used a U-net derived 1-D CNN based on the envelopes (Homomorphic, Hilbert, Wavelet, PSD) of the PCG signal for the detection of the heart sound events. They found that the use of larger 1D kernels leads to better results because more time steps can be used for the classification of a time step. Their model achieved an average sensitivity of 93.9% and an average positive predictive value of 94% in detecting $S_1$ and $S_2$ sounds. Yin et al. (18) used a TCN, which combines the advantages of time information processing of RNNs and the incorporation of expressive local information of CNNs. They added a Viterbi algorithm as the state annotation algorithm to their TCN architecture to ensure the physiological fixed state sequence rule ($S_1$, systole, $S_2$, diastole), achieving an F1 score of 97.02%. Fernando et al. (19) proposed a Bidirectional Long Short-Term Memory architecture with a soft attention mechanism, achieving a competitive F1 score of 96.70% with only 17k parameters for the heart sound events $S_1$ and $S_2$.

### C Lung event detection

**1) Model architectures:** Most of the works on the topic of diagnostic support for cardiac and pulmonary auscultation focus on the classification of the audio files into either healthy or pathological (20–26), the classification of normal breathing sounds, and various types of adventitious sounds (26–39). There are few works that have made attempts to determine the sound events precisely in time with start and end time using strong labels. In SED of lung sounds, the recent works benchmarked RNNs, CNNs and CRNNs (40). Hsu et al. (40) trained eight different model architectures for each sound event separately and showed that CRNN Bidirectional Gated Recurrent Unit models were most performant in detecting the sound events inhalation, exhalation, Continuous Adventitious Sounds (CAS), and Discontinuous Adventitious Sound (DAS) events (40,41). Messner et al. also used a BiGRU model in (42) for simultaneously predicting breathing phases (inhalation/exhalation) and the presence/absence of crackles on a non-public internal dataset, obtaining an average event-based F1 score of ≈ 72% for crackles and of ≈ 86% for breathing phase events. In (43), Hsiao et al. used an attention-based encoder-decoder model for breath-phase detection and achieved an F1 score of ≈ 90% for inhalation and of ≈ 93% for exhalation detection in 100 ms time windows on a non-public internal dataset. Jácome et al. (44) trained a faster region-based CNN (Faster R-CNN) model on a non-public internal dataset (3212 inspiratory events, 2,842 expiratory events). The model achieved a sensitivity of 97.5% and specificity of 85.0% for inspiration detection and a sensitivity of 95.5% and specificity of 82.5% for expiration detection.

At the time of writing, there are no known works to identify polyphonic, i.e. temporally simultaneous, events using lung or heart auscultation databases. The main known work on lung sound event recognition based on the published HF_V1 (40) dataset has trained and separately evaluated individual models for each sound event in the experiments performed. Thus, no comparative values are available for the test accuracy of lung SED by a single model. A clinical-practical use, such as for real-time monitoring purposes of the heart and lungs, requires strict latency requirements that preclude the parallel use of multiple models for signal processing. With this work, we investigate for the first time how a single model can be trained with competitive accuracy to cover such applications in the future.

## III. METHODOLOGY

### A. Lung sound datasets

*1) HF_Lung_V1 database (40):* This dataset is with 9765 15-second audio files from 279 subjects the largest open-access lung sound database available. It contains 34095 inhalation, 18349 exhalation, 8457 wheezing, 4740 rhonchi, 686 stridor and 15606 crackle labels. Audio recordings were collected from two different study cohorts and with two different auscultation devices, namely Littmann 3200 (3M, Saint Paul, Minnesota, USA), and HF_Type-1 (Heroic Faith Medical Science, Taipei, Taiwan). Different from the authors' experiments of HF_Lung_V1, where separate models were trained for each sound event, we trained a single model for polyphonic SED with an output of (*n* classes *x* timesteps) using the existing training-test split (7809 recordings for training and 1956 for testing). For the validation split, we used 10% of the training data.

*2) LEOSound dataset:* The LEOSound dataset consists of 482704 recordings using the commercially available respiratory sound monitoring device LEOSound (Löwenstein Medical GmbH & Co. KG, Bad Ems, Germany) to record lung sounds. The auscultations were collected overnight between 22:00h to 06:00h from different subgroups of different ages. The night-time sleep monitoring auscultations are split into 30-second recordings and were collected continuously by three adhesive microphones attached dorsally over the right and the left lung and laterally over the trachea of the patients. This dataset has no information or labels other than clinical information about the patients. Using two specialized single-task heart and lung SED models (see Fig. 3), we generated 18 191 736 PL for $S_1$, 18205784 PL for S2, 4095807 PL for inspiration, 3987145 PL for expiration, 536956 PL for crackles (DAS), 298884 PL for wheezes and 181385 PL for rhonchi, totaling 45497697 PL.

### B. Heart sound datasets

*1) Physionet / CinC 2016 (Training set of the challenge):*
The dataset was assembled by merging different smaller cardiac auscultation databases with great differences in the recording environments, recording sensors, recording durations, and recording quality. It is structured in six databases (from a to f) and contains a total of 3153 recordings from 764 subjects. All recordings are commonly resampled to 2kHz, and durations vary between 5 and 120 seconds. The dataset provides strong labels for the four heart cycle sound events $S_1$-systole-$S_2$-diastole, which were first generated by an LR-HSMM-based segmentation algorithm and afterwards manually corrected. We included only $S_1$ and $S_2$ events because systole and diastole are derived from these. Noisy parts - which are labeled by "(N" and "N)" - were excluded.

*2) CirCor DigiScope Phonocardiogram dataset:*
The CirCor dataset is the largest cardiac auscultation dataset with a total of 5272 audio recordings. Auscultations were collected from a primarily pediatric study population ranging between 0 and 21 years of age; recording duration varies from 4.8 to 80.4 seconds. Similar to the PhysioNet/CinC 2016 dataset, cardiac event annotations were initially generated by three SED models, which were then reviewed and corrected. The dataset was published on the occasion of the George B. Moody PhysioNet Challenge 2022 (45). At the time of this work, only 60% of the total dataset was available for use as a training set in version 1.0.3, as the remaining 40% was withheld as a test set for the challenge.

### C. Manual test set annotation
To quantify the performance of background signal annotation in addition to foreground signal scores, validation ground truth labels were created for the two main datasets, CirCor (46) and HF_V1 (40). Thereby, audio recordings were audio-visually analysed for the presence of heart and lung sounds by a trained physician and an experienced investigator. Each auscultation sound recording was labeled by only one researcher. Regular consensus meetings were arranged between the researchers to evaluate the labeling of the sound events and to reach a consensus in case of differences. In heart sound auscultations, we annotated the lung sound events inspiration and expiration; in lung sound auscultations, we annotated the heart sound events $S_1$ and $S_2$.

### D. Pre-processing, Feature Extraction

The input feature for all SED models is a log Mel spectrogram with 64 Mel bins, extracted directly, without any filtering, from a 10-second-long audio data window with a sampling rate of 4kHz, without any normalization applied. The strong label information (onset and offset times) has been converted to a multi-hot vector of $\{0,1\}^{N \times M}$ where $N$ and $M$ are the number of time frames in a sound clip and the number of sound event classes, respectively. Inactive time steps are marked with 0, and active time steps with 1, if one of the event classes ($S_1$, $S_2$, inspiration, expiration, wheezes, crackles, rhonchi, stridor) is active.

### E. Training strategy

Our semi-supervised learning approach follows three steps (see Fig. 3): i) We first trained a baseline CNN-BiGRU model with the ground truth labels of the foreground signal of the cardiac and lung auscultation databases, ii) used the trained models to generate the respective pseudo-label for the background signals, and then iii) trained a new model on all data, using the combination of ground truth labels and pseudo-labeled auscultation files (see Fig. 3). We evaluated the corresponding models separately with respect to the benchmark database, comparing the test accuracy of the trained models in the respective categories. We injected noise and made use of data augmentation in the wave domain (gain, high-pass, low-pass, white noise, pitch shift, reverb, time dropout, noise injection) and in the spectrogram domain (height/frequency stretching, spec augment (47), filter augment (48)) during training. In a last step, we used the trained model from the third step (iii) to generate a total of 45497697 pseudo-labels for foreground and background signals of a much larger corpus of unlabeled audio files (482704 recordings) and used this data in a new training iteration. We also created a separate test data set with lung sound labels for the CirCor cardiac auscultation database, (47) and with heart sound labels for the HF_V1 lung auscultation database (40) to quantify the test accuracy of the detection of the respective background signal.

### F. Training challenges and filtering

Several problems for training a unified heart and lung SED model using pseudo-labels and ground truth labels arise from, i) the domain invariance (stethoscope/sensor, auscultation site, environment, patients), ii) the class imbalance, iii) the severe data imbalance between active and inactive frames of the different event classes. To ensure that an actual signal was present during pseudo-label generation, we only integrated the pseudo-labels including the audio recording, if this resulted in possible vital signs according to physiological measures. In the case of heart event detection, we integrated pseudo-labels only if the resulting heart rate between 40 and 240; for lung auscultation, we chose a respiratory rate range between 0 and 35. We further used the Asymmetric Focal Loss function proposed by Imoto et al. (13) in our experiments to address class and active/inactive frame imbalance.

### G. Model architectures

**1) CNN-BiGRU**: For the baseline network architecture, we used CNN-BiGRU, which is widely used as a baseline SED system, such as in DCASE2018 challenge task (49). Fig. 4 shows that the model consists of three convolutional blocks. Each convolutional block contains two convolutional layers (kernel sizes of 3 × 3). Batch normalization (50) and ReLU non-linearity (51) is used after each convolutional layer. 2 × 1 average pooling is applied after each convolutional block for downsampling in the frequency domain. In Fig 4., the number following „@" represents the number of feature maps. The 3-dimensional outputs of the last convolutional block are averaged over the frequency axis, yielding a 2-dimensional audio feature representation which is then provided as input to the first 256-dimensional bidirectional GRU layer. Then, after the recurrent layers, a time distributed fully connected layer with sigmoid non-linearity is applied to predict the presence probability of sound events for each time step. The output layer has one neuron for each class. We additionally applied unsupervised pretraining of the convolutional feature extraction layers by using BYOL-A (52) and transformed the weights to the CNN layer block for the SED task.

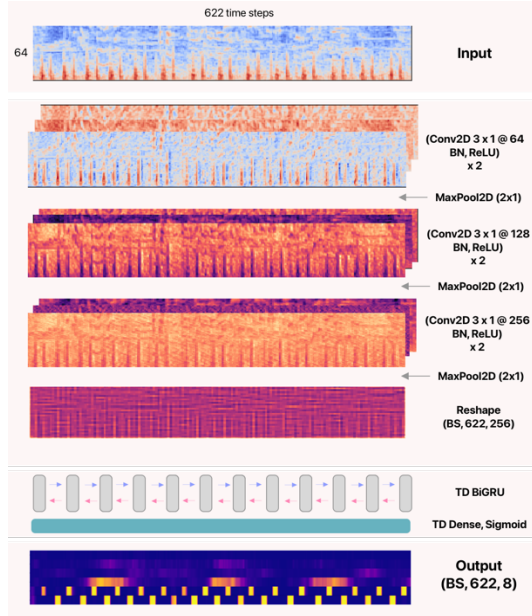

**Figure 4:** *Used convolutional bi-directional recurrent neural network (CRNN) architecture for polyphonic SED. Images are extracted from the model and upscaled for visualization purposes. The CNN layers are adapted from PANN (53) and pretrained using BYOL-A (52). TD = Time Distributed, BS = Batch Size, BN = Batch Normalization.*

*2) Temporal convolutions*

As a second model, we used the compute efficient SELD-TCN framework proposed by Guirguis et al. (54), which showed superior performance to CRNN architectures in SED tasks, while being up to 40 times faster at model inference on GPUs. We have adapted the model architecture according to the specified input and increased the convolutional filter count to 256.

## *H. Model training*

Models were implemented using TensorFlow (version 2.10.0) and CUDA (11.2). Training was performed with two NVIDIA RTX A6000 GPUs. During training, we randomly selected a 10-second window (622 frames) from an audio recording. The model output is a 3-dimensional tensor of shape (batch size, time steps, $n$ classes). Models were trained for 25 epochs using binary cross-entropy (BCE) loss and Asymmetric Focal Loss (AFL) (13,55,56) with $\gamma = 0.0625$ and $\zeta = 1.0$ as the objective and the Adam (57) optimizer, using mini-batch of size 64, and a learning rate of $10^{-4}$. Early stopping is used to reduce over-fitting. The output of each prediction for a 10-second audio window was a sequential prediction matrix of size 622 (time steps) × 8 (classes). At test time we threshold model predictions by setting values > 0.5 to 1 (active) and the rest of the values to 0 (inactive).

## VI. EXPERIMENTS

## *A. Evaluation metrics*

We calculated event-based metrics and segment-based metrics to evaluate the experiments using the *sed_eval* package (58) at the recording level. Apart from threshold binarization, no post-processing was applied before evaluation. Performance was assessed using the F1 score as it is better suited for imbalanced dataset evaluations.

*1) Event-/collar-based metrics*

We considered sound events to be correctly detected when the temporal position overlapped with that of a ground truth event of the same class within a temporal tolerance (time collar) (see Fig. 5, D). Following previous literature protocols, we set a tolerance value of ± 60ms for the evaluation of heart SED, and of ± 500ms (42) for lung SED, because heart and lung sound events have strongly different event timespans. We defined the following at the recording level:

- **True positive:** Events, in which the distance between the determined position (prediction) and the target position (GT) is equal or smaller than the tolerance window and thus overlap.
- **True negative:** Ground truth and predicted labels both indicate an event to be inactive.
- **False positives:** The event from the prediction is not present in in the ground truth labels.
- **False negative:** In the ground truth labels an event occurs which is not present / not recognized in the prediction.

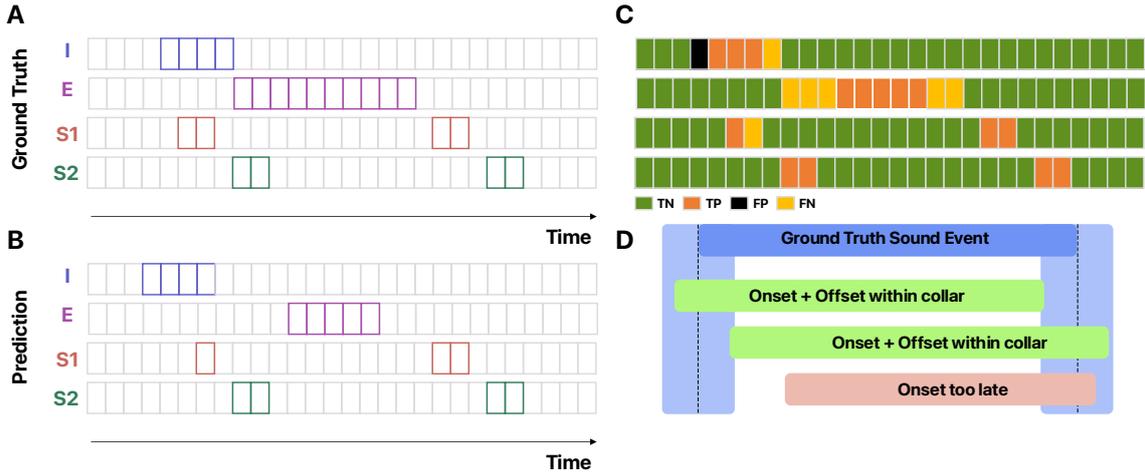

**Figure 5:** *Examples of ground truth (A) and prediction labels (B) at given time steps for a segment-based evaluation (C) and an event-based evaluation (D). In (D), the t collar time window is shown in a lighter shade of blue than the ground truth sound event, and the dashed line marks the center/half of the t collar window for the onset and offset time points, respectively. Inactive time steps are displayed in grey, active sound events are shown in color in (A) and (B). Graphic is adapted from (58). I = inspiration, E = expiration, S1 = first heart sound, S2 = second heart sound, TN = true negative, TP = true positive, FP = false positive, FN = false negative*

*2) Segment-based*

We first used the onset and offset times of the ground truth event labels to create the ground truth time segments, creating vectors of $\{0, 1\}^{N \times M}$, where $N$ is the number of time frames in a recoding and $M$ is the number of sound event classes (see Fig. 5 A and B). Next, by comparing the ground truth time steps (active bars are coloured, inactive are displayed in grey in Fig. 5 A) with the results prediction (bars in Fig. 5 B), we defined time-step wise:

- **True positive** (orange bars in Fig. 5 C),
- **True negative** (green bars in Fig. 5 C),
- **False positive** (black bars in Fig. 5 C),
- **False negative** (yellow bars in Fig. 5 C)

Afterwards, the F1 scores for classifying the individual time steps in each audio recording was determined.

*3) Lung SED metrics*

Additionally, we used the evaluation scheme for the lung sound events introduced by the authors of the HF_V1 lung database. They used an extensional evaluation protocol based on the similarity of events to the Jaccard Index (JI) (40). Following the authors' definition (40), the JI was used to evaluate the extent to which a detected event overlapped with a ground truth event:

- **True positive**: JI value greater than 0.5,
- **False negative**: JI value was between 0 and 0.5
- **False positive**: JI value was 0

*4) Heart sound detection metrics*

The $S_1$ (1st heart sound) and $S_2$ (2nd heart sound) events were used to determine the F1 scores of the localization of the heart sounds. Systole, by definition, is between the first two heart sounds and diastole begins at the end of the second heart sound and is bounded by the onset of the next first heart sound. Previous published work has used different tolerance window (time collar) values for event-based metrics when evaluating the validity of the onset and offset times. Springer used a time collar value of 100ms (16), Schmidt used 60 ms (15), and Messner used 40 ms (17). To make an even-handed comparison to prior work, we chose a time collar value of 60 ms.

## B. Results
*1) Performance evaluation in comparison with baselines*

To our best knowledge, at the time of this writing, there are two other works (19,40) that have trained and evaluated SED models using the HF_V1 lung auscultation database. However, the authors have trained separate models for each sound event class. The event classes stridor, rhonchi, and wheezes were further combined as CAS by the authors of the HF_V1 database and thus a total of four models were evaluated separately. We are not aware of any work that has used the HF_V1 dataset to create a model for detecting multiple (overlapping) lung sound event classes. For this reason, we first trained a multi-class base model for the six sound event classes (inspiration, expiration, wheezes, crackles, stridor, rhonchi), and subsequently trained models for simultaneous detection of heart and lung sound events with the goal of not regressing with respect to the base model. Because Hsu et al. combined the lung sound events wheezes, stridor, and rhonchi as CAS, we also treated them as one event at evaluation time. The results of all trained models for the HF Lung V1 and Physionet 2016 datasets are provided in Table I.

## 2) Baseline model scores

Using the evaluation scheme of (40), our best-performing base model achieved an F1 score of 61.2% for inspiration, 34.8% for expiration, 22.6% for crackles, and 23.6% for CAS (see Table I). In comparison to the baseline results, our heart and lung SED model regressed slightly in heart sound detection with F1 scores of 94.2% for $S_1$ and 92.7% for $S_2$ and it gained significantly in breath phase detection with relative improvements of 11.6% for inspiration, 5.1% for expiration, while losing 8.0% for crackles, and 2.5% for CAS.

**Table I:**
Experimental results for the benchmark databases HF_V1 and Physionet. Since different working groups used different evaluation metrics for the two benchmark datasets, we used the authors' evaluation protocol in the case of HF_V1 (JIE) and additionally report T-collar-based event metrics (TBE). For the evaluation of cardiac tone segmentation of the Physionet/CinC 2016 database, we specified T-collar-based event metrics (TBE) and segment metrics (Seg). Blue-shaded fields indicate areas for which predictions cannot be generated with the models. Bold values correspond to the best value achieved for a given metric.

| | Model | HF V1 Dataset ($n$ = 1873, 24320.94 sec, t_collar 500 ms) | | | | | | | | | | | | Physionet ($n$ = 568, 12734.14 sec, t_collar 60 ms) | | | |
|---|---|---|---|---|---|---|---|---|---|---|---|---|---|---|---|---|---|
| | | I F1 score | | | E F1 score | | | DAS F1 score | | | CAS F1 score | | | $S_1$ F1 score | | $S_2$ F1 score | |
| | | TBE | Seg. | JIE | TBE | Seg | JIE | TBE | Seg | JIE | TBE | Seg | JIE | TBE | JIE | TBE | JIE |
| | | | | | | | | Baseline LS | | | | | | | | | |
| - | **CRNN w/ BCE** | 66.4 | 69.1 | 61.2 | 38.9 | 46.1 | 34.8 | **28.3** | 32.3 | 22.6 | **31.1** | 31.3 | 23.6 | / | / | / | / |
| - | TCN w/ BCE | 64.0 | 72.3 | 59.0 | 35.6 | 42.0 | 28.0 | 11.6 | 10.4 | 05.1 | 18.8 | 17.1 | 12.2 | / | / | / | / |
| | | | | | | | | Baseline HS | | | | | | | | | |
| - | CRNN w/ BCE | / | / | / | / | / | / | / | / | / | / | / | / | 95.0 | 96.9 | 93.6 | 94.3 |
| - | TCN w/ BCE | / | / | / | / | / | / | / | / | / | / | / | / | 94.6 | 96.5 | 93.2 | 94.1 |
| | | | | | | | | Proposed Multi-Task Models | | | | | | | | | |
| - | CRNN w/ BCE | 72.5 | 76.8 | 68.7 | 40.8 | 48.4 | 35.8 | 25.8 | 29.6 | 18.8 | 15.2 | 15.0 | 08.8 | 94.0 | 96.0 | 92.2 | 93.0 |
| - | TCN w/ BCE | 71.2 | 77.7 | 66.7 | 39.0 | 45.2 | 29.2 | 5.4 | 05.1 | 02.1 | 07.1 | 07.2 | 04.1 | 94.0 | 96.0 | 92.4 | 93.0 |
| - | **CRNN w/ AFL** | **72.8** | 77.8 | 69.2 | **43.3** | **49.6** | **39.9** | 18.8 | 24.5 | 14.6 | 26.7 | 27.0 | 21.1 | 94.2 | **96.1** | 92.7 | 93.4 |
| - | TCN w/ AFL | 69.3 | 77.1 | 64.1 | 37.8 | 47.9 | 30.4 | 13.2 | 15.2 | 6.6 | 18.9 | 18.6 | 12.3 | 93.6 | 95.6 | 92.2 | 93.1 |
| † | CRNN w/ BCE | 71.9 | 77.6 | 68.2 | 39.8 | 45.7 | 35.8 | 26.8 | 33.2 | 22.7 | 24.2 | 28.4 | 17.8 | 94.0 | 96.0 | 92.2 | 93.0 |
| † | TCN w/ BCE | 72.1 | 77.7 | 67.3 | 37.0 | 45.8 | 29.5 | 17.8 | 17.8 | 09.2 | 03.9 | 05.5 | 03.2 | 94.0 | 96.0 | 92.4 | 93.0 |
| † | **CRNN w/ AFL** | 72.6 | **78.0** | 69.2 | 42.5 | 48.8 | 38.2 | 28.0 | **32.8** | **24.6** | 30.5 | **32.9** | **23.7** | 94.2 | 96.0 | 92.7 | 93.4 |
| † | TCN w/ AFL | 67.9 | 74.7 | 62.7 | 34.3 | 39.2 | 25.5 | 12.6 | 07.7 | 04.3 | 01.4 | 01.8 | 0.05 | 93.6 | 96.0 | 92.2 | 93.1 |

† Extra Data: 45M pseudo-labeled events from 482.704 audio recordings

## 3) Evaluating the effect of extra data

As depicted in Table I all models trained with extra data (LEOSound dataset, values are marked with the symbol † in Table I) achieved higher F1 scores for the lung sound events DAS and CAS compared to those without, while those of the events inspiration and expiration remained approximately the same. Our proposed CRNN model with AFL achieved relative improvements for DAS of 5.0% and of 0.1% for CAS. At the same time, the accuracy of heart event detection did not decrease and remained approximately stable at 94.2% for $S_1$ and 92.7% for $S_2$.

## 4) Evaluation of class-wise performance

The F1 scores of the different lung sound event classes showed substantial differences. For a more detailed examination of the individual sound event classes, it is helpful to compare the precision-recall curves for different threshold values between 0 and 1. Figure 6 shows the sound event-wise precision-recall curves of all sound classes with our proposed best-performing CRNN model. The blue and red curves represent the test and validation precision-recall curves, respectively. From Fig 6, it can be observed that some sound event classes, such as inspiration, have high precision at a variety of thresholds. It can be also seen that the validation and the evaluation precision-recall curves have similar trends, but do not overlap, indicating that the distribution of validation and evaluation data differs, especially for the expiration and rhonchi events. Fig. 6 also indicates that different sound classes have different thresholds for achieving optimal metrics, such as the F1 score.

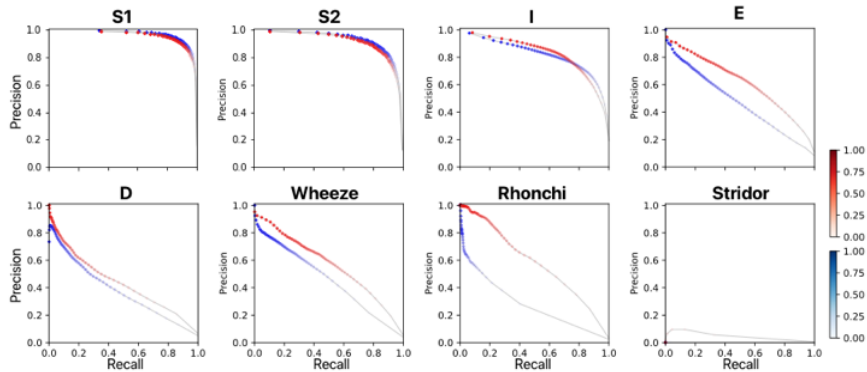

**Figure 6:** *Precision-recall curves of sound events at different thresholds for $S_1$, $S_2$, inhalation, exhalation, D (discontinuous adventitious sounds, = crackles), and CAS (= wheeze + rhonchi + stridor) detection. The blue and red curve show the test and validation precision-recall curve, respectively. The graphic is adapted from (59).*

This observation is also confirmed in Fig. 7, which shows the mean absolute percentage error (MAPE) values at various thresholds between 0 and 1. The blue and red curves represent the test and validation precision-recall curve, respectively. It can be observed that the optimal threshold for sound events with high test accuracy is around 0.5 ($S_1$, $S_2$ and I). In these cases, the MAPE curves have a similar flat trend. For the sound events with lower test accuracies, such as D, wheeze, and rhonchi, the MAPE curves are less symmetrical. Here, higher accuracies are achieved at threshold values below 0.5. For the stridor sound event, the MAPE value reaches 1 from a threshold value greater than 0.5.

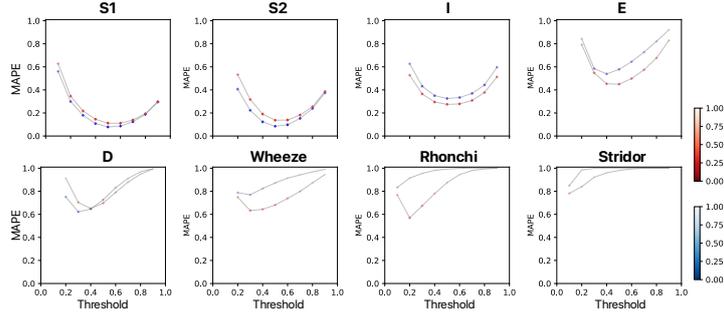

**Figure 7:** *Mean absolute percentage error (MAPE) curves for S1, S2, inhalation, exhalation, D (discontinuous adventitious sounds, = crackles), and wheeze, rhonchi, and stridor event detection. Lower values indicate better performance. The blue and red curve show the test and validation precision-recall curve, respectively. The graphic is adapted from (59).*

### 5) Evaluating multi-class multi-task models vs. single-class

Table II displays the current state-of-the-art results for F1 scores of single-class SED for HF_V1 in comparison to our proposed multi-class SED models. By comparing the results obtained with a multiple model inference setup (one model per sound event class) with those obtained with only one model, the clear drop of the F1 test scores becomes apparent, especially in the underrepresented classes expiration, DAS, and CAS.

**TABLE II**
Segment-wise & JI-based F1 scores of the HF Lung V1 dataset for the detection of Inhalation (I), Exhalation (E), Continuous Adventitious Sounds (CAS), Discontinuous Adventitious Sounds (DAS) events.

|  | I | | E | | DAS | | CAS | |
|---|---|---|---|---|---|---|---|---|
|  | Seg | JIE | Seg | JIE | Seg | JIE | Seg | JIE |
| CNN-BiGRU (40) | **80.1** | 85.9 | 62.4 | 68.4 | 69.9 | **69.5** | 52.6 | **51.6** |
| TCN (19) | n/a | **94.2** | n/a | **83.5** | n/a | **73.7** | n/a | **88.0** |
| OURS | 78.0 | 69.2 | 49.6 | 39.9 | 32.8 | 24.6 | 32.9 | 23.7 |

### 6) Comparison with state-of-the-art Heart Sound Detection models

Previous work on heart SED has used slightly different evaluation schemes. Consistent with previous sound event detection challenges, we report the F1 scores determined by event-based metrics for the two events $S_1$ and $S_2$ (time collar = 60ms). The comparison of results is shown in Table III. We achieved competitive results with both our baseline results and with the best performing heart and lung event detection model. In particular, the multi-task lung and heart sound event detection model underwent only a marginal relative loss of 0.8% for $S_1$ and 0.9% for $S_2$ compared with the specialized heart sound event detection model.

**TABLE III**
Event-based F1 scores of the Physionet/CinC 2016 dataset for the detection of $S_1$ and $S_2$ events.

|  |  | $S_1$ | $S_2$ |
|---|---|---|---|
| Author | Model | F1 | F1 |
| Messner (17) | BiRNN | 97.5 | 95.0 |
| OURS | CRNN (specialized) | 95.0 | 93.6 |
| OURS | CRNN (multi-task) | 94.2 | 92.7 |

### 7) Performance background signals

To demonstrate that the combined training of heart and lung sound event labels leads to better recognition of background sound events, we first performed an evaluation on a separately created subset of CirCor in which we manually labeled the lung sounds inspiration ($n$ = 799) and expiration ($n$ = 87) in 212 files. The event-based results with a time collar of 500ms showed that our model achieved comparably high F1 scores with 70.5% for inspiration and 42.5% for expiration as the obtained F1 scores in the HF_V1 (lung sound foreground) evaluation. Similarly, to evaluate the trained models for the detection of (background) heart sounds in lung auscultations, we labeled $S_1$ and $S_2$ heart sounds in a randomized subset of H1_V1 (test split). In comparison between the model trained using only the foreground heart sound signal, the trained model on foreground GT labels and background pseudo-labels achieved relative improvements for $S_1$ of 3.6%, and 3.0% of S2 (time collar 60 ms).

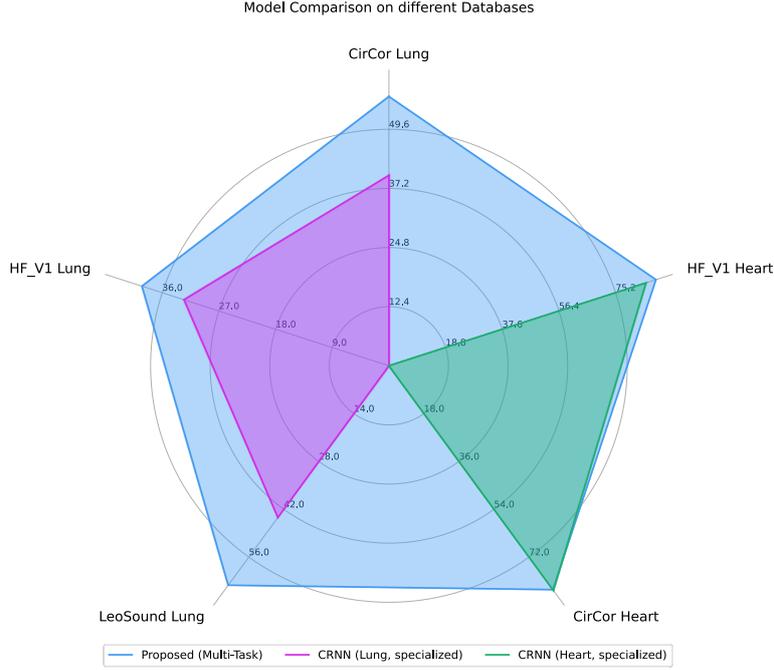

**Figure 8:** *Comparison of our proposed multi-task SED model with the other baseline single-task specialized models the different databases CirCor Lung (background HS signal), HF_V1 Heart (background HS signal), CirCor Heart (foreground HS signal), LEOSound Lung (foreground LS signal), HF_V1 Lung (foreground LS signal). Depicted are the class-wise macro-average F1 scores for each task and dataset.*

Fig. 8 summarizes the performance on the established key benchmarks for lung and heart SED of our proposed multi-task multi-class model compared to the single-task (specialized) baseline models. While in the case of heart SED, the baseline model achieved higher test accuracies (foreground heart signal, CirCor dataset), the baseline models underperformed our model in all other categories. The combined training scheme of heart and lung auscultation databases had a slightly positive effect on background heart sound detection in the HF_V1 Heart subset. In contrast, the test accuracies for lung auscultation turned out to be much higher for CirCor Lung, HF_V1 Lung, and LEOSound Lung.

## VII. Discussion

This work has investigated a new semi-supervised training strategy whereby SED models are trained to simultaneously detect heart and lung sound events by merging previously separate datasets for the first time. We first trained separate SED models using the public challenge datasets for heart and lung SED, and then used these to enrich the databases with pseudo-labels for the background sound events. Although the pseudo-labels are likely to contain a lot of noise and error, by merging the ground truth and pseudo-labels we were able to train multi-task SED models that did not regress in lung SED in comparison to the single-task baseline models and only slightly regressed in heart SED. We were also able to show, using manually created labels, that our multi-task SED models achieved high test F1 scores in detecting background sound events, i.e., detecting heart sounds in the case of dorsal lung auscultation or detecting lung sounds in the case of thoracic heart auscultation. We observed that performances were inconsistent across the heavily imbalanced lung sound event categories, which might be due to class and active/inactive frame imbalance during training. The state-of-the-art single-class models in (40) and (19), which are not affected by class imbalance between the sound event classes, achieve more balanced test accuracies. Yet, it is not practical to train separate SED models for each of the eight individual sound event classes for real-time inference scenarios due to latency and compute requirements. Similarly, it must be emphasized that for anatomical-physiological reasons, it will not be possible in many cases to capture all heart and lung sounds with only one auscultation point. Especially in the case of strictly unilateral or regional pathologies, only the proven strategy of anatomically assigned auscultation points will be diagnostically valid, e.g. in the case of lobar pneumonia or pneumothorax. This means that model optimization will never solve the issue of a missing or insufficient signal.

## VIII. CONCLUSION

In this work, we propose a single common model for detecting heart and lung events, regardless of the auscultation point. With our proposed training strategy, we have trained multi-class multi-task SED models for a single-stage inference to obtain heart and lung sound events. We achieved competitive results for the detection of the first two heart sounds and for inspiration. Our results showed that by merging different databases in a training iteration it is feasible to train a SED model for the simultaneous detection of heart and lung sounds. To our knowledge, this is the first report on the multi-task challenge of detecting heart and lung sound events independent of the site of auscultation. To better understand the implications of these results and to achieve more balanced accuracies between sound event classes, new strategies besides AFL should be investigated. In a broader context, our findings could lead the way to i) affordable self-instructed auscultation for patients especially in remote locations and ii) a decision support system to ease practitioners' daily challenges